\documentclass{ragtime} 
\title[Ultrafast outflows from MAD discs]%
      {MAD UFOs: Magnetically Arrested Discs with persistent Ultra-Fast Outflows}

\author[P. Sukov\'{a}, 
        M. Zaja\v{c}ek
        and V. Karas]
       {Petra Sukov\'{a}\at{1,a} 
        Michal Zaja\v{c}ek\at[]{2} 
        and Vladim\'ir Karas\at[]{1}\\
        \ins{1}Astronomical Institute of the Czech Academy of Sciences, Bo\v{c}n\'{\i} II 1401,\splitins[1]  
        CZ-180\,00 Prague, Czech Republic
        \\
        \ins{2}Department of Theoretical Physics and Astrophysics, Faculty of Science,\splitins[1]  Masaryk University, Kotl\'a\v{r}sk\'a 2, CZ-611\,37 Brno, Czech Republic\\
        \ins{a}\Email{petra.sukova@asu.cas.cz}} 

\newcommand{\apjl}{Astrophys. J. Lett.} 
\newcommand{\apjs}{Astrophys. J. Suppl. Ser.} 
\newcommand{\ao}{Appl. Opt.} 
\newcommand{\aap}{Astron. Astrophys.} 

\begin{document}

\begin{abstract}
General-relativistic magneto-hydrodynamical (GRMHD) simulations of accreting black holes suggest that the accretion flows form toroidal structures embedded in a large scale component of magnetic field, which becomes organized on length-scales exceeding the gravitational radius of the central black hole. Magnetic field grows gradually until a Magnetically Arrested Disc (MAD) develops that diminishes or inhibits further accretion. We study an outflow that develops in the MAD state in 3D GRMHD simulations. We show that the outflow can be accelerated to relativistic velocities and persist over the course of our simulation. We compare the properties of the outflow from MAD discs with those launched by orbiting secondary at close orbit. The main difference is that the orbiting body launches a more coherent, quasiperiodic ultrafast outflow at lower velocities ($v<0.5c$) while the outflow launched in the MAD state (without the body) has a stochastic behaviour and has an approximately flat velocity distribution between lower anf higher outflow velocities, $0.2c<v<0.3c$ and $v>0.5c$.  
\end{abstract}


\section{Introduction}\label{intro}
In the course of the recent decade, high-resolution data brought compelling evidence that the electromagnetic variability of accreting black holes has its origin in the hot plasma orbiting on event-horizon scales 
\citep{2018A&A...618L..10G,2019ApJ...875L...1E}.
Despite these observational breakthroughs, the precise factors that determine the accretion rate as well as the origin of the outflows, flares, and oscillations seen in these systems remain largely unknown. With the upcoming missions aiming to observe black hole systems at ever-increasing sensitivity and resolution
(e.g. LSST, Athena, ESO Extremely Large Telescope), an improvement in understanding of these dynamical processes is of prime importance. 

In a recent study \citep{nas-stellar-transits}, we focused on the observable effects that can be caused by repetitive transits of perturbers through accretion flows near supermassive black holes (SMBHs).  
We have modified the GRMHD {\tt HARMPI} code \citep{2015MNRAS.454.1848R} 
 to assume a rigid body moving along geodesics in a hot accretion flow. 
 An important outcome of this study is the existence of quasiperiodic ultrafast outflows from the central region launched by the transit of the perturber through the accretion disc. 
 The body penetrates through the disc and kicks off blobs of magnetized plasma into the empty, but strongly magnetized funnel region. 
 This material moves at a mildly relativistic speed $\sim (0.1$--$0.5)c$ along the boundary between the funnel and the accretion torus. One of the possible observational effects could be quasiperiodic absorption events if the blobs cross our line of sight and the partially ionized gas can absorb a fraction of the X-ray emission of the underlying accretion flow. 
 Moreover, when the body is large enough and sufficiently close to the SMBH, the accretion rate can be significantly influenced by the close fly-by, yielding periodic changes to the luminosity and the spectral shape of the source. 
 
Recently, a previously quiescent galactic nucleus in a nearby galaxy ($z=0.056$) went into outburst and regular monitoring of the source revealed absorption events with quasiperiodically changing column density \citep{ASSASN}. These absorption events can be interpreted as an outflow with the velocity of $\sim 0.3c$, whose column density varies by approximately one order of magnitude with a period of 8.5 days. Due to the fact, that the achieved velocity is about one order of magnitude larger than the usual stellar wind, such highly ionized absorbers are called an ultra-fast outflow \citep[UFO;][]{2003MNRAS.345..705P,2010A&A...521A..57T}.
Under an assumption of one absorption event per orbit and in combination with the mass estimate of the central SMBH, $\log(M/M_\odot) = 7.5 \pm 0.5$, this gives us the characteristic orbital distance of the perturber $r_{\rm per}\sim 90M$ \footnote{We work in the geometrized unit system with $G=c=1$, hence velocities become dimensionless and times and lengths can be measured in terms of $M$ (using standard SI or cgs units, gravitational radii are defined as $r_g = GM/c^2$ and gravitational times as $GM/c^3$).}.
Several possible scenarios explaining these observations are discussed in detail in \citet{ASSASN}, showing that the perturber-induced outflow is a viable mechanism to produce such an outcome. If such an observational pattern could be unambiguously classified as a large mass-ratio binary system, high-cadence, large field-of-view X-ray surveys will help to identify more candidates for close binaries in active galactic nuclei (AGN). 
   
Binaries, in which at least one of the components is an SMBH are the main extragalactic targets for the upcoming European space-based GW observatory LISA  \citep{2016PhRvD..93b4003K,2017PhRvD..95j3012B}. 
The targeted binaries can have various objects as the other component, ranging from stellar-origin black holes or neutron stars (SOBHs/NSs) through intermediate-mass black holes (IMBH), to other SMBHs. 
In all cases, the GW emission causes a gradual degradation of the orbit 
up to the merger. If and when such events are seen by LISA, it will be of the utmost scientific interest to localize the host galaxy, which would make it a multi-messenger standard siren for cosmology \citep{holz2005using,2016JCAP...04..002T}. 

In this paper, we will focus on the competing possible mechanism causing the quasiperiodic UFOs, in which the UFOs originate in the magnetic reconnection events in the innermost region of the so-called magnetically arrested accretion disc (MAD disc). By means of GRMHD simulations, we will compare the properties of the outflow launched by the motion of the perturber and that emanating from the temporal variability of the MAD disc stemming from the interchange instability of the plasma near the horizon. We will show that even though the 2D simulations of MAD discs yield strong periodic flares with larger amplitudes in the accretion and outflow rates, relaxing the axial symmetry assumption in 3D simulations leads to much less fierce evolution with rather stochastic variability. Moreover, the velocity of the outflow originating in the inner $\sim10$--$20$ gravitational radii achieves higher values than observed in the above-mentioned system.

In summary, the accretion in the MAD disc state does not seem to explain the observed features of the UFO better than the perturber-induced-outflow scenario. 

This contribution is structured as follows. In Section~\ref{numerics}, we introduce the numerical setup of the GRMHD simulations. The results are presented in Section~\ref{results} and we conclude with Section~\ref{conclus}.

\section{Numerical setup}\label{numerics}

We study the behavior of the plasma numerically using the general relativistic magnetohydrodynamical code HARMPI \citep{2015MNRAS.454.1848R,2007MNRAS.379..469T}, which is based on the HARM code \citep{0004-637X-589-1-444}.
The code, which we have equipped with our own modifications, is 3D
and parallelized. The perturber is moving along a geodesic trajectory in Kerr spacetime, assuming the accreting gas has a negligible gravitational field with respect to the central SMBH. The effect of the gravitational wave losses or the hydrodynamical drag on the secondary body can be omitted because they are negligible in such a setup during the time span of our simulation (covering tens to hundreds of orbits). Its dynamical effect on the gas is simulated assuming that all gas within the sphere of influence of the orbiting body characterized by the radius $\mathcal{R}$ is comoving with the body. 
The details of the numerical procedure are given in \citet{nas-stellar-transits}.

The simulations proceed in two steps. In the first step, we spontaneously evolve the initial accretion torus, which is obtained as the exact solution of a magnetized torus with a non-constant angular momentum profile in equilibrium \citep{Witzany_Jefremov-tori}. We omit the equilibrium toroidal magnetic field, instead, we equip the torus with poloidal magnetic loops, so that the MRI starts working shortly after the beginning of the simulation and the torus starts to accrete. After an initial transient time, a quasi-stationary accretion state is achieved. 

\begin{figure}[htbp]
\begin{center}

\includegraphics[width=\linewidth]{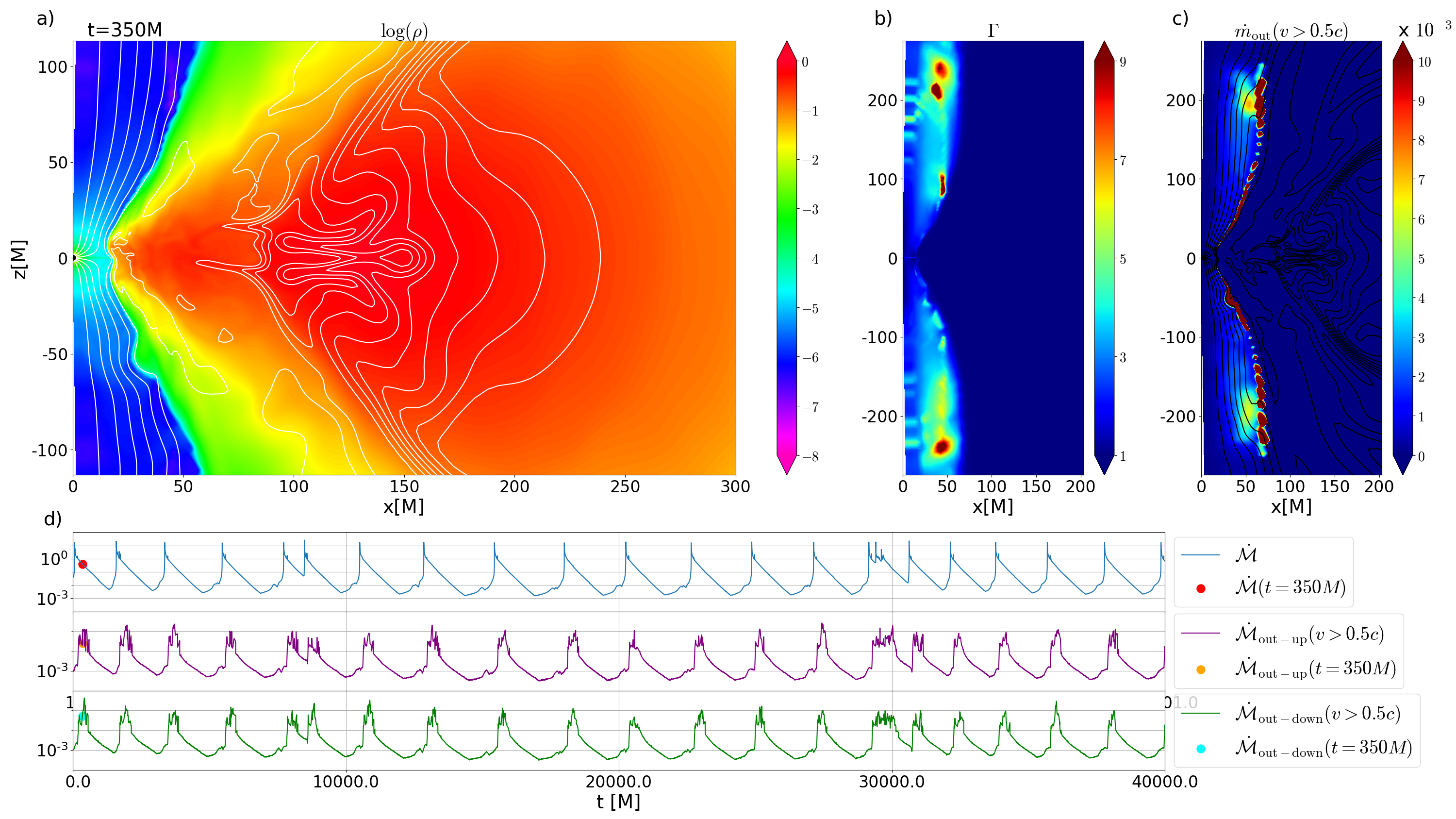}
\end{center}
\caption{\label{Fig:2D-MAD-slice}An instant from the 2D GRMHD numerical simulation of the accretion flow in the MAD state. The density, the Lorentz factor, and the outflowing rate density 
is shown in the first row, the time dependence of the accretion rate (blue), and the outflowing rate directed upward (purple) and downward (green) are shown in the bottom panel. In the two-dimensional simulation, the pressure of the accumulated organized large-scale vertical magnetic field balances the gas pressure, the accretion is halted and proceeds only intermittently via the interchange instability close to the horizon, which results in the simultaneous accretion event and ejection of relativistic outflow to both directions. }
\end{figure}

We use two different configurations of the initial magnetic field. First, the magnetic field lines follow the isocontours of the density of the torus, forming one big magnetic loop over the whole torus. Such initial conditions are known to be prone to quick formation of the MAD state, in which the magnetic pressure close to the horizon balances the gas pressure, the accretion is shut down and proceeds only via the interchange instability when small blobs of matter accrete onto the black hole during the reconnection of the magnetic field lines \citep{1974Ap&SS..28...45B,1976Ap&SS..42..401B,2003PASJ...55L..69N}.
The other initial condition consists of several smaller loops with alternating polarity, which leads to a so-called SANE (standard-and-normal evolution) state.

After achieving the quasi-stationary accretion state (see Fig.~\ref{Fig:2D-MAD-slice}), we restart the simulation. At that point, we can regrid our simulation domain from 2D to 3D in such a way, that we copy the 2D slice into all $\phi$ directions, introducing a $\sim 1\%$ random perturbation in density to relax the axial symmetry and enable the evolution in the $\phi$-direction (axial symmetry is preserved by the numerical scheme). At this point, we can also turn on the motion of the perturber.

With two 3D GRMHD simulations, we aim to show the distinctive nature of the outflow launched by the magnetic reconnection yielding the magnetic flux eruptions in the MAD accretion state (run 3D-MAD) and the outflow driven by the transiting body through the accretion flow (run 3D-star). We also provide long-term evolution of the 2D-MAD simulation for comparison (see Table~\ref{Tab:runs} for the overview of the runs). 

\begin{figure}[htbp]
\begin{center}

\includegraphics[width=0.9\linewidth]{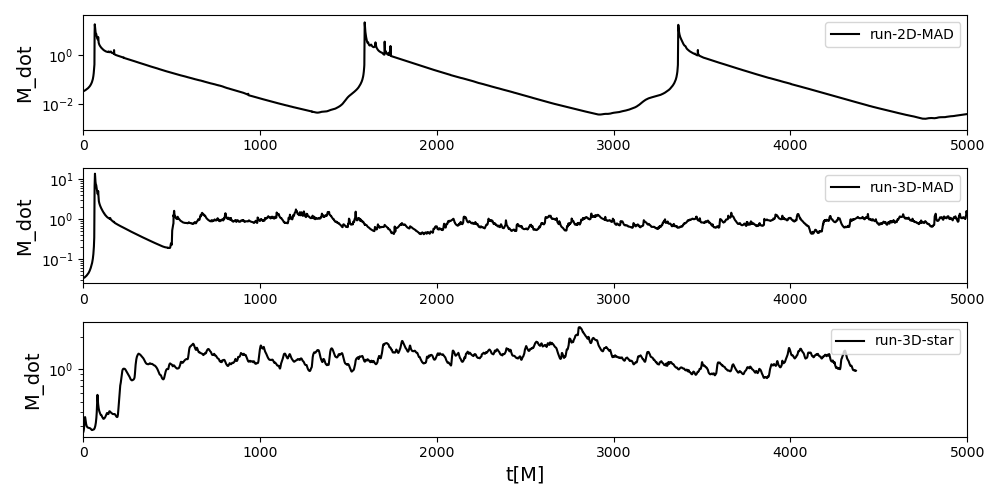}
\end{center}
\caption{\label{Fig:mdots}Time dependence of the accretion rate for all three runs shown in the time interval $[0,5\,000]M$. }
\end{figure}

\section{Results}\label{results}

For our study of the MAD state, we have chosen one slice from a 2D simulation of a MAD accretion torus. At first, we continue this 2D simulation for $40\,000M$ (run-2D-MAD). In the first row of Fig.~\ref{Fig:2D-MAD-slice}, we show panels with the density in logarithmic scale, the Lorentz factor, and the outflow with $v>0.5c$. In the bottom, the three panels show the time profile of accretion rate $\dot{\mathcal{M}}$ and outflowing rate $\dot{\mathcal{M}}_{\rm out} (v>0.5c)$ directed into the upper or lower part of the funnel. We can see the time-correlated large-amplitude quasi-periodic flares in accretion rate and outflows upwards and downwards persisting during the whole simulation.

\begin{table}[bp]
\begin{center}
\begin{tabular}{ccccccc} 
\hline
run & $a$ & resolution
  & $t_{\rm max}$ & $\mathcal{R}[M]$ & $i$ & $e$\\\hline
run-2D-MAD & 0.9 & 384 x 192 & $40\,000M$ & --- & ---\\\hline
run-3D-MAD & 0.9 & 384 x 192 x 96 & $5\,000M$ & --- & ---\\\hline
run-3D-star & 0.5 & 384 x 256 x 96 & $4\,370M$ & 2 & $65.5^\circ$ & 0.19\\\hline
\end{tabular}
\caption{\label{Tab:runs} Table of the performed GRMHD runs. In the columns, we list the name of the run, the SMBH spin $a$, the resolution of the grid, the total duration of the run, and in case of binary simulation, the influence radius $\mathcal{R}$ of the secondary and the inclination and eccentricity of the orbit.}
\end{center}
\end{table}

\begin{figure}[htbp]
\begin{center}

{\scriptsize 3D-MAD \hspace{5.2cm} 3D-star}

\includegraphics[width=0.49\linewidth]{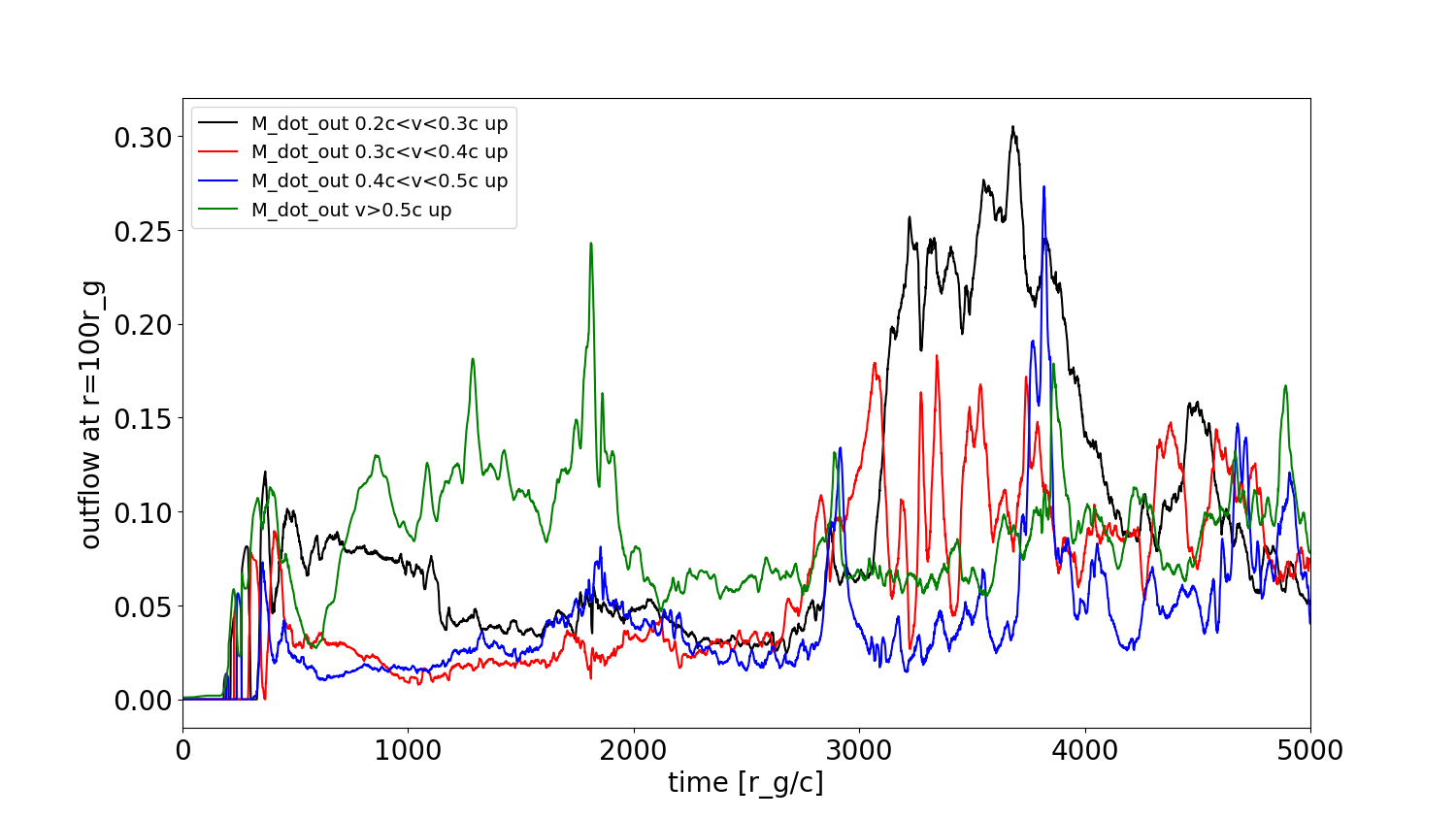}
\includegraphics[width=0.49\linewidth]{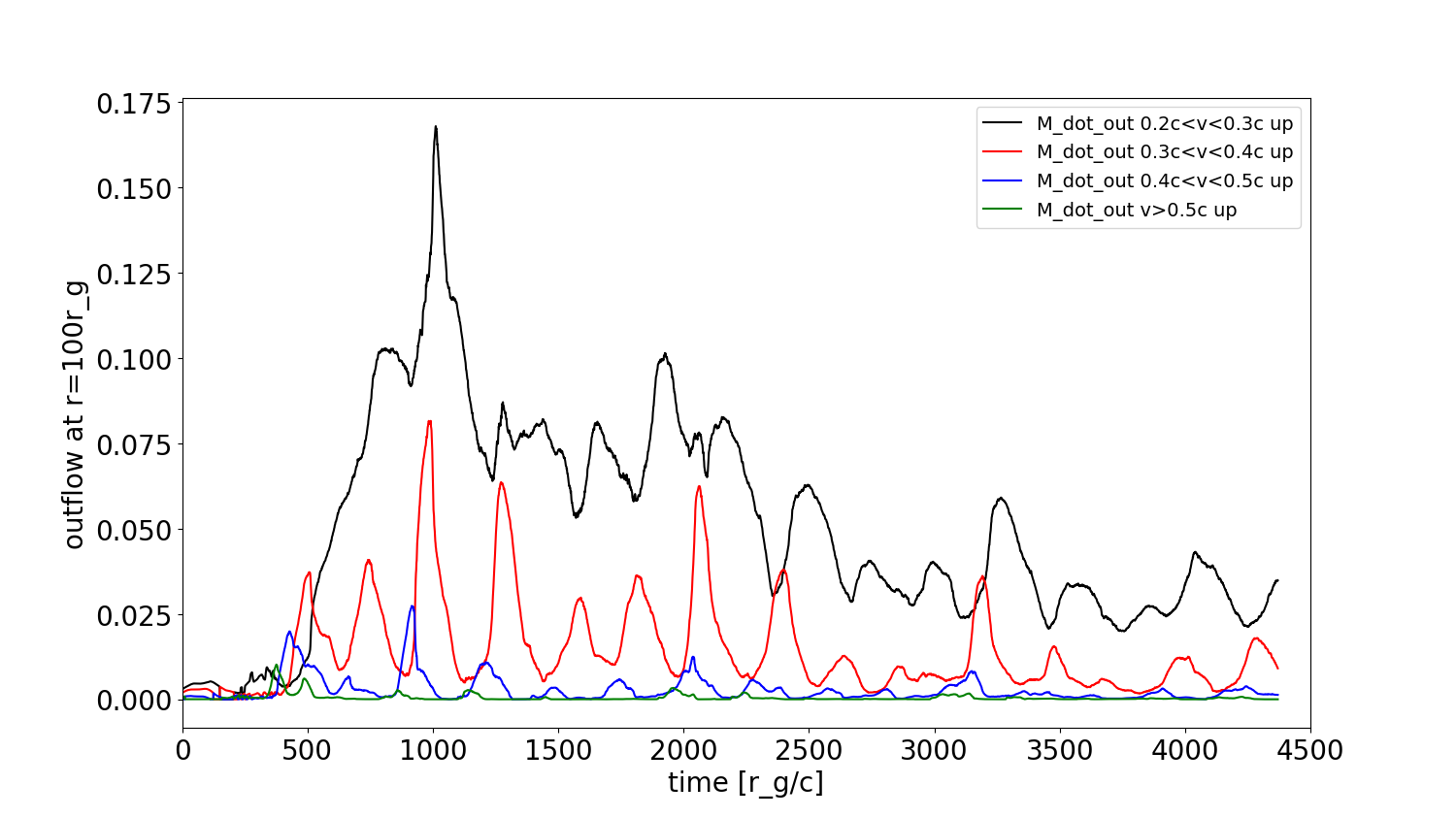}

\includegraphics[width=0.49\linewidth]{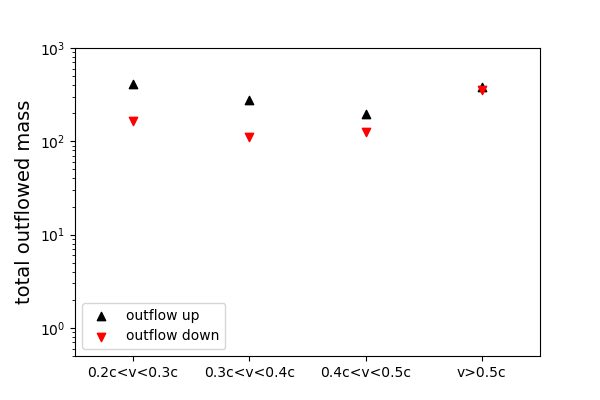}
\includegraphics[width=0.49\linewidth]{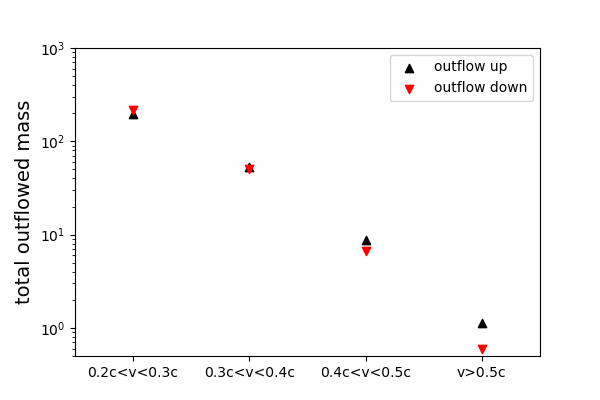}
\end{center}
\caption{\label{Fig:outflows} In the first row, we show the time profiles of the ultrafast nuclear outflow measured at $r=100M$. On the left, we show the case corresponding to run-3D-MAD, while on the right, we show run-3D-star. With different colors, the outflow rates in 4 different velocity bins are depicted. The second row shows the total outflow in the upward (black) and downward (red) directions during the simulation in the four bins. }
\end{figure}

\begin{figure}[tbp]
\begin{center}
\includegraphics[width=0.49\linewidth]{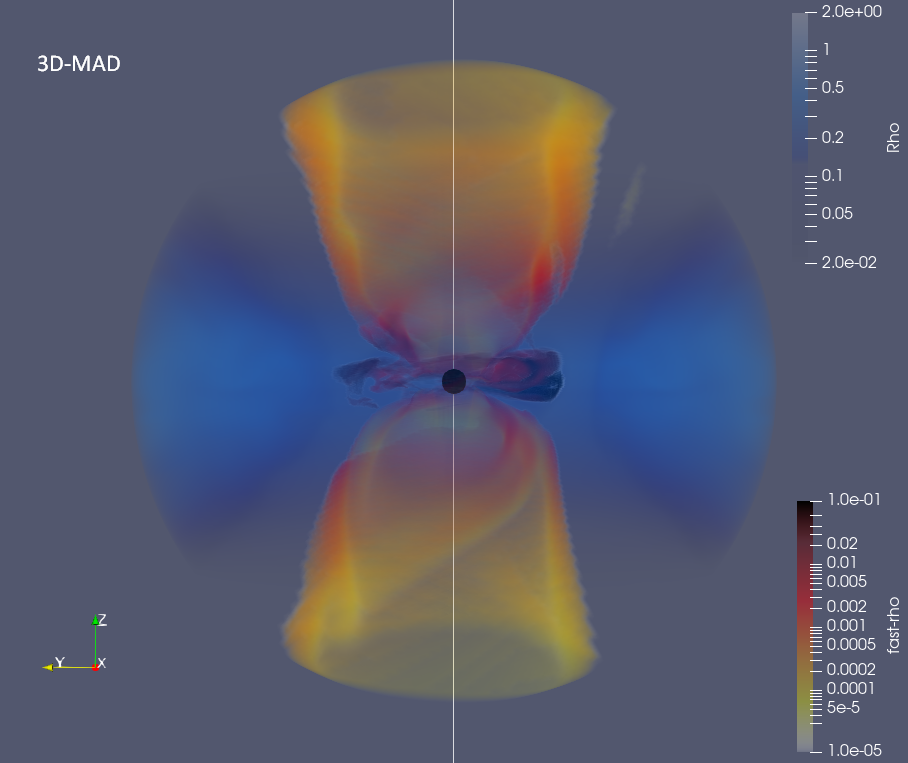}
\includegraphics[width=0.49\linewidth]{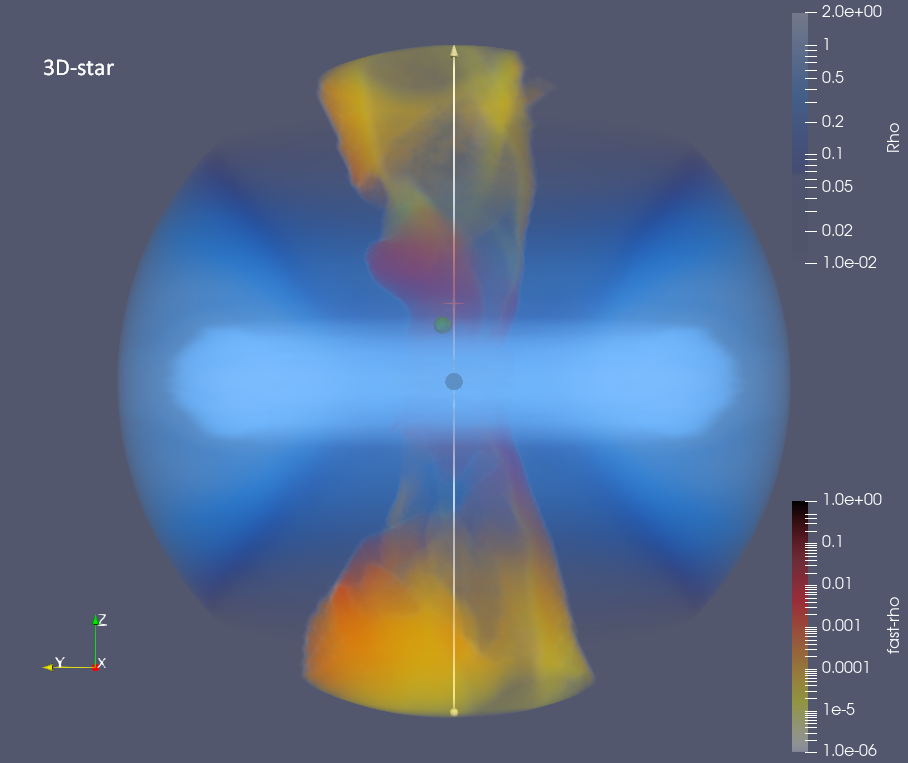}
\end{center}
\caption{\label{Fig:3D-snapshots}Snapshots from 3D simulations showing the slowly moving gas (with $\gamma < \gamma_{\rm thr}$ in the blue color scale) and fast outflowing gas (with $\gamma > \gamma_{\rm thr}$  and positive radial velocity) in the yellow-red-black color scale. On the left, we show run-3D-MAD with $\gamma_{\rm thr}=1.05$, while on the right, we show run-3D-star with $\gamma_{\rm thr}=1.02$.}
\end{figure}

Starting with the same initial conditions as run-2D-MAD we initialize a 3D run (run-3D-MAD) with a total duration of $5\,000M$. 
In Fig.~\ref{Fig:mdots}, we show the comparison of the accretion rate for all the runs. We can notice that in the 3D simulation, the disc still completes one cycle similar to the ones seen in the 2D simulation (due to our regrid procedure), but after the axisymmetry is completely lost, the accretion proceeds in a much more steady regime, the large peaks and dips are smoothed. This is due to the fact that in 3D, the individual gaseous blobs are accreting from different directions, following the magnetic reconnection events, therefore, the averaged accretion rate is more steady and does not show large-amplitude flares. Instead, smaller blobs are expelled during the accretion, which are getting stretched in the azimuthal direction, forming long dense filaments on a helical trajectory (see the left panel in Fig.~\ref{Fig:3D-snapshots}). This finding is in agreement with a recent particle-in-cell simulation of reconnection-driven flares by \citet{AaA202346781}. 
During the duration of run-3D-MAD, the accretion rate or the outflowing rates do not show prominent quasi-periodicity. However, to probe the periodicity on the time scales of about $2\,000M$ (reported e.g. in  \citet{10.1093/mnras/stac3330}) up to about $6\,000M$, which corresponds to the observed variability, a longer 3D run is needed, which is currently beyond the scope of this contribution.

In run-3D-star, due to the computational cost of 3D simulations, we need to put the orbiting perturber on a tighter orbit than the derived characteristic distance $r_{\rm pert}=90M$ for the above-mentioned source so that we are able to cover several consecutive revolutions of the body. The orbit is inclined with $i=65.5^\circ$ from the equatorial plane and slightly eccentric ($e=0.19$), with pericenter $r_p=10M$ and apocenter $r_a = 14.68M$. The fundamental periods in $r,\theta,\phi$ directions \citep{Fujita_2009} are $P_r=370M, P_\theta=273M, P_\phi=270M$ (i.e. about 20x shorter than the measured period). Hence, the total duration of the simulation ($4370M$) covers about 16 orbits of the body, which is in line with the observed 13 peaks during the outburst of the aforementioned galactic nucleus \citep{ASSASN}. 

While the accretion rate, shown in Fig.~\ref{Fig:mdots}, does not show strong periodicity correlated with the motion of the perturber, when looking at the corresponding time profiles of the outflowing rates, we do notice a qualitative difference. In Fig.~\ref{Fig:outflows} we show the outflow rates evaluated at $r=100M$ in 4 different velocity bins, i.e. $0.2c<v<0.3c$, $0.3c<v<0.4c$, $0.4c<v<0.5c$, $v>0.5c$. We can see prominent quasiperiodicity corresponding to the motion of the perturber for the outflow velocity in the range of $0.2c<v<0.5c$.
Furthermore, while the total outflowing mass in each velocity bin is rapidly decreasing for the run-3D-star, in the case of the MAD accretion flow, the outflow with $v>0.5c$ is larger than the outflow in three lower-velocity bins (see Fig.~\ref{Fig:outflows}, second row). As a result, a much higher amount of mass with $v>0.3c$ is expelled than in the perturber-induced outflow model, which can serve as a distinguishing factor between the MAD-driven nuclear outflows and the perturber-driven outflows.

\section{Conclusions}\label{conclus}

We have performed two 3D GRMHD simulations, which correspond to the advection-dominated, hot accretion flow in the MAD state and the one perturbed by an orbiting body. The simulations have shown the different nature of the outflow induced by the magnetic reconnection events near the horizon in the MAD state and the outflow expelled by the transits of the secondary body through the accretion flow. The main difference lies in the more coherent periodicity of the outflows caused by the orbiting perturber on the one hand and the higher speed reached by the outflow launched by the MAD disc on the other hand.

\ack

PS has been supported by the fellowship Lumina Quaeruntur No.\ LQ100032102 of the Czech Academy of Sciences. This work was supported by the Ministry of Education, Youth and Sports of the Czech Republic through the e-INFRA CZ (ID:90140). VK and MZ acknowledge the Czech Science Foundation project (ref.\ GF23-04053L).


\bibliography{ragsamp}
\end{document}